\documentclass[twocolumn,showpacs,amsmath,pre]{revtex4}
\usepackage{graphicx}
\usepackage{amsmath}
\usepackage{amssymb}
\usepackage{natbib}
\usepackage{subfigure}
\usepackage{color} 

\begin{document}

\title{Particle-scale reversibility in athermal particulate media below jamming}

\author{Carl F. Schreck$^{1,2}$}
\author{Robert S. Hoy$^{3}$}
\author{Mark D. Shattuck$^{4,1}$} 
\author{Corey S. O'Hern$^{1,2,5}$} 

\affiliation{$^{1}$Department of Mechanical Engineering \& Materials Science, 
Yale University, New Haven, Connecticut 06520-8260, USA}

\affiliation{$^{2}$Department of Physics, Yale University, New Haven,
Connecticut 06520-8120, USA}

\affiliation{$^{3}$Department of Physics, University of South Florida,
Tampa, Florida 33620, USA}

\affiliation{$^{4}$Benjamin Levich Institute and Physics Department, The City 
College of the City University of New York, New York, New York 10031, USA}

\affiliation{$^{5}$Department of Applied Physics, Yale University, New Haven,
Connecticut 06520-8120, USA}

\begin{abstract}
We perform numerical simulations of athermal repulsive frictionless
disks and spheres in two and three spatial dimensions undergoing
cyclic quasi-static simple shear to investigate particle-scale
reversible motion.  We identify three classes of steady-state dynamics
as a function of packing fraction $\phi$ and maximum strain amplitude
per cycle $\gamma_{\rm max}$.  Point-reversible states, where
particles do not collide and exactly retrace their intra-cycle
trajectories, occur at low $\phi$ and $\gamma_{\rm max}$.  Particles
in loop-reversible states undergo numerous collisions and execute
complex trajectories, but return to their initial positions at the end
of each cycle. Loop-reversible dynamics represents a novel form of
self-organization that enables reliable preparation of configurations
with specified structural and mechanical properties over a broad range
of $\phi$ from contact percolation to jamming onset at $\phi_J$. For
sufficiently large $\phi$ and $\gamma_{\rm max}$, systems display
irreversible dynamics with nonzero self-diffusion. 
\end{abstract}

\pacs{
83.80.Fg
61.43.-j,
63.50.Lm,
62.20.-x
}

\maketitle

In equilibrium thermal systems, reversible processes are imagined as
transitions from one thermodynamic microstate to another with no
change in free energy. For example, thermal fluctuations in
high-temperature fluids give rise to particle motions that yield only
very small changes in entropy.  However, finite deformations of
supercooled liquids and amorphous solids, upon reversal of the strain,
can produce microscopically {\it irreversible} motion, such as
collective particle rearrangements~\cite{schall}, anelasticity, and
plastic flow~\cite{bmg}.  The identification of topological defects in
crystalline materials is straightforward, whereas it is much more
difficult to identify particle-scale motion that gives rise to
plasticity in amorphous materials~\cite{stz}.

Granular materials, foams, and other athermal particulate media are
highly dissipative, and therefore must be driven to induce particle
motion.  Experimental studies of granular media have shown macroscale
reversibility of the packing fraction during cyclic
shear~\cite{nicolas} and vibration~\cite{nowak}.  Since these systems
are far from thermal equilibrium, one might assume that they do not
display {\it microscale} reversible motion when subjected to cyclic driving.
Experimental and computational studies of 2D foams have
identified both reversible and irreversible T1 bubble neighbor
switching events during cyclic shear~\cite{Lundberg2008}.  Researchers
have also shown that motion of individual particles transitions from
reversible to irreversible beyond a density-dependent critical strain,
which decreases with increasing packing fraction, in cyclically sheared
suspensions~\cite{Pine2005,Corte2008}.

An important open question is whether athermal particulate media can
undergo completely reversible motion due to inter-grain collisions
when subjected to cyclic loading. We address this question by
performing numerical simulations of frictionless granular materials in
two and three spatial dimensions undergoing quasistatic cyclic simple
shear~\cite{slotterback} over a wide range of packing fraction $\phi$
and shear strain amplitude $\gamma_{\rm max}$.  We identify two
classes of grain-scale reversible motion, point and loop (which are
stable to finite perturbations).  For point-reversible dynamics,
particles do not collide during the forward cycle, and thus they
exactly retrace their trajectories upon reversal. In contrast,
particle collisions occur frequently during loop-reversible dynamics,
but the system self-organizes so that particles return to the same
positions at the beginning of each cycle, despite complex particle
motion during the cycle.  We map out the `dynamical phase diagram' as
a function of $\phi$ and ${\gamma}_{\rm max}$.  The system transitions
from point- to loop-reversible and then from loop-reversible to
irreversible (with nonzero self-diffusion) dynamics with increasing
$\phi$ and $\gamma_{\max}$. We show that the time evolution toward
steady-state point- and loop-reversible behavior can be collapsed onto
a universal scaling function with power-law scaling at short and
intermediate times, and exponential decay at long times.

We perform numerical studies of $N$ athermal disks
and spheres undergoing quasi-static, cyclic simple shear in 2D
and 3D at constant $\phi$ using shear-periodic 
boundary conditions in square or cubic cells~\cite{allen}. 
Particles interact via the pairwise, purely repulsive linear spring
potential
\begin{eqnarray}
V(r_{ij}) = \frac{\epsilon}{2} \bigg(1-\frac{r_{ij}}{\sigma_{ij}}\bigg)^2 \Theta(\sigma_{ij}-r_{ij}),
\end{eqnarray}
where $r_{ij}$ is the center-to-center separation between particles
$i$ and $j$, $\Theta(x)$ is the Heaviside step function, $\sigma_{ij}
= (\sigma_i+\sigma_j)/2$, and $\sigma_i$ is the diameter of particle
$i$.  We focus on bidisperse particle size distributions, {\it i.e.}
$50$-$50$ mixtures by number with diameter ratio
$\sigma_l/\sigma_s=1.4$, to frustrate crystallization during
shear~\cite{gao}. We consider system sizes from $N=32$ to $512$ to
assess finite size effects for packing fractions below and near the
onset of jamming ($\phi_J \sim 0.84$~\cite{shen} in 2D and $\sim
0.65$~\cite{xu} in 3D).

The particles are initially placed randomly in the simulation
cell at packing fraction $\phi$ and then relaxed using
conjugate gradient energy minimization~\cite{gao}.  We 
apply simple shear strain by  
shifting each particle horizontally 
\begin{eqnarray}
x_{n,k+1}^i = x_{n,k}^i + \Delta\gamma y_{n,k}^i,
\label{shear_strain}
\end{eqnarray}
in increments of $\Delta\gamma=10^{-3}$, where $x_{n,k}^i$ and
$y_{n,k}^i$ are coordinates of particle $i$ at the $k$th step of
strain cycle $n$~\cite{supp}. After each strain step, we minimize the total
potential energy at fixed shear strain, $\gamma_k=k \Delta \gamma$ for
the forward or $\gamma_k = 2 \gamma_{\rm max} - k\Delta \gamma$ for
the reverse part of the cycle.  This process is repeated
for up to $n=10^6$ cycles.

During the simulations, we measure the single cycle mean-square
displacement (at step $k=0$ for each cycle)
\begin{eqnarray}
\Delta r^2_1(n) & = & (N \sigma_s^2)^{-1} \sum_i \Big(
(X_{n,0}^i-X_{n+1,0}^i)^2 \nonumber \\
& + & (Y_{n,0}^i-Y_{n+1,0}^i)^2+
(Z_{n,0}^i-Z_{n+1,0}^i)^2 \Big)
\label{msd}
\end{eqnarray}
and arc length
\begin{eqnarray}
L^2(n) & = & (N\sigma_s^2)^{-1} \sum_i 
\bigg( \sum_k \big[
(X_{n,k+1}^i-X_{n,k}^i)^2 \\ 
& + & 
(Y_{n,k+1}^i-Y_{n,k}^i)^2 + (Z_{n,k+1}^i-Z_{n,k}^i)^2 \big]^{1/2} \bigg)^2 \nonumber
\label{arc_length}
\end{eqnarray}
versus $n$, where $X_{n,k}^i = x_{n,k}^i-\gamma_ky_{n,k}$, $Y_{n,k}^i
= y_{n,k}^i$, and $Z_{n,k}^i = z_{n,k}^i$ are the non-affine
displacements of particle $i$ after subtracting off the affine
contribution.  The long-time dynamics are either reversible or
irreversible depending on $\phi$ and $\gamma_{\rm max}$.  We quantify
the steady-state behavior by measuring $\Delta r_1^2(n)$ and $L^2(n)$
of the intra-cycle particle trajectories.

\begin{figure}[h]
\begin{center}
\includegraphics[width=0.45\textwidth]{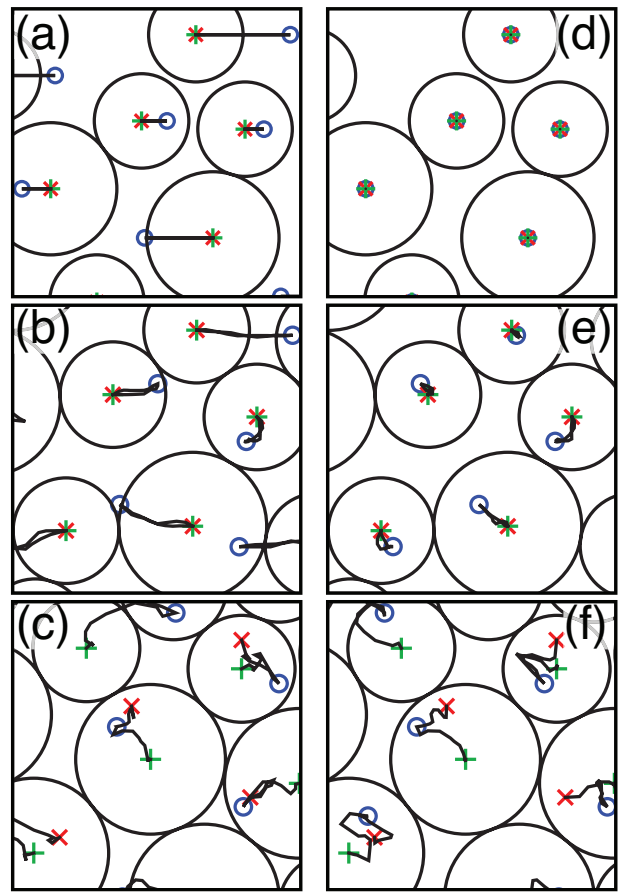}
\caption{(color online) Particle tracks (solid lines) in a small
window of a $N=128$ system of bidisperse disks undergoing (a)
point-reversible, (b) loop-reversible, and (c) irreversible behavior
during cyclic simple shear. Panels (d), (e), and (f) show the disks'
tracks in panels (a), (b), and (c), respectively, after subtracting
off the affine motion.  The systems in (a) and (d) correspond to
$\phi=0.64$ and $\gamma_{\rm max}=0.8$, (b) and (e) to $\phi=0.8$ and
$\gamma_{\rm max}=0.8$, (c) and (f) to $\phi=0.82$ and $\gamma_{\rm
max}=0.8$. Pluses, circles, and crosses mark the beginning, middle,
and end of the particle tracks, and particle outlines correspond to
the beginning of the cycle.}
\label{Fig1}
\end{center}
\end{figure}

Particles in point-reversible systems organize to avoid collisions. At
long times, no collisions take place, and $L(n) = \Delta r_1(n) = 0$,
or more aptly, they fall below small numerical thresholds, {\it e.g.}
$\Delta r_1(n) < \tau_r = 5 \times 10^{-4}$ and $L(n) < \tau_L =
10^{-8}$.  The values of
$\tau_r$ and $\tau_L$ do not qualitatively affect our results as long
as they are sufficiently small.  Particle motions for point-reversible
systems are affine and in the direction of the imposed affine shear
(Fig.~\ref{Fig1} (a)). Thus, the non-affine tracks of each particle
are zero (Fig.~\ref{Fig1} (d)).

\begin{figure}[h]
\begin{center}
\includegraphics[width=0.45\textwidth]{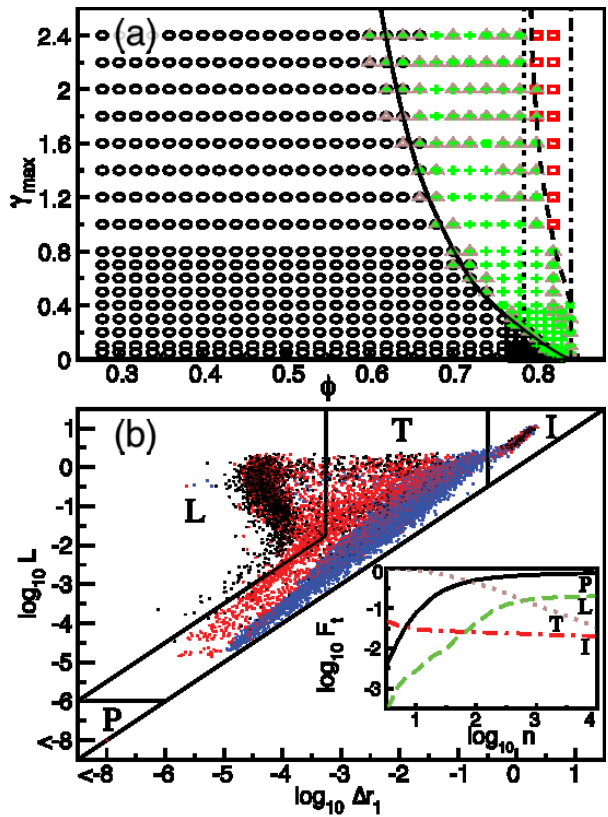}
\caption{ (color online) (a) ``Dynamical phase diagram'' for $N=128$
bidisperse disks after $n=10^4$ cycles showing point-reversible
(circles), period-one (pluses) or multi-period (crosses)
loop-reversible, and transient (triangles) or steady-state (squares)
irreversible dynamics versus $\gamma_{\rm max}$ and $\phi$. The solid
and dashed lines indicate $\phi_L(\gamma_{\rm max})$ and $\phi_{\rm
I}(\gamma_{\rm max})$, the boundaries between point- and
loop-reversible dynamics and between loop-reversible and irreversible
dynamics, respectively. The vertical dotted and dot-dashed lines
define $\phi_R = 0.785$ and $\phi_J \approx 0.84$. (b) Intra-cycle
mean-square displacement $\Delta r_1$ versus arc-length $L$ after
$n=10$ (blue), $3\times10^{2}$ (red), and $10^4$ (black) cycles.  The
regions labeled $P$, $L$, $T$, and $I$ define point-reversible,
loop-reversible, transient irreversible, and steady-state irreversible
dynamics, respectively. The $P$-region extends from $L = \Delta r_1=
10^{-6}$ to $10^{-16}$ (not shown). The dashed boundaries indicate
$\Delta r_1 = \tau_r$, $\Delta r_1 = 0.3 \sigma_s$, $L=\tau_L$, $L =
30 \Delta r_1$, and $L=\Delta r_1$ discussed in the main text.  The
inset shows the fraction $F_t$ of systems in (a) categorized as
point-reversible (solid), loop-reversible (dashed), transient
irreversible (dotted), and irreversible (dot-dashed) dynamics versus
$n$. Similar data for 3D systems is shown in Supplementary Material.}
\label{Fig23}
\end{center}
\end{figure}

In {\it loop-reversible} systems, particle collisions occur
frequently, but the system self-organizes so that particles return to
the same positions at the start of each cycle. Since collisions
between particles occur, $L(n) > 0$, but $\Delta r_1(n) = 0$ ({\it
i.e.}  below $\tau_r$). Particle trajectories (Fig.~\ref{Fig1} (b))
and non-affine displacements (Fig.~\ref{Fig1} (e)) trace out complex
paths, yet all particles end up in the same locations at the beginning
of each new cycle, {\it i.e.}  $X_{n+1,0}^i=X_{n,0}^i$ and
$Y_{n+1,0}^i=Y_{n,0}^i$ in 2D.  Thus, particle trajectories form
closed loops in configuration space. We focus on period one
loop-reversible systems, but multi-period dynamics are also
found. Particle trajectories are elongated in the direction of affine
displacement, whereas nonaffine displacements are more compact.

Particles in systems undergoing {\it irreversible} dynamics do not
return to their original positions at the beginning of each new cycle,
{\it e.g.} $X_{n+1,0}^i\ne X_{n,0}^i$ and $Y_{n+1,0}^i\ne Y_{n,0}^i$
in 2D.  (Figs.~\ref{Fig1} (c) and (f).) Irreversible systems have
nonzero $\Delta r_1(n)$ and $L(n)$ ({\it i.e.} $L(n)>\tau_L$ and
$\Delta r_1(n)>\tau_r$).  Systems can be ``transient'' irreversible in
time and evolve into point- or loop-reversible systems, or
steady-state irreversible and remain irreversible in the large-cycle
limit with nonzero self-diffusion.

The steady-state ``dynamical phase diagram'' in Fig.~\ref{Fig23} (a)
for cyclically sheared athermal disks shows point- and
loop-reversible, as well as irreversible regimes versus $\phi$ and
$\gamma_{\rm max}$~\cite{supp}. Point-reversible systems occur at low
$\phi$ and $\gamma_{\rm max}$, whereas irreversible systems occur for
$\phi \gtrsim \phi_J$.  At intermediate packing fractions between
roughly contact percolation~\cite{shen} and jamming onset, {\it e.g.}
$0.6 \lesssim \phi \lesssim 0.84$ in 2D, loop-reversible systems are
found.  The boundary between point- and loop-reversible systems is
$\gamma_{\rm max} \sim A(\phi) (\phi_J - \phi)^{\lambda} \Theta(\phi_J
- \phi)$, where $A(\phi)$ depends weakly on $\phi$ and $\lambda \sim
1.2 \pm 0.1$ for $\phi \rightarrow \phi_J$ and $2.2 \pm 0.2$ for $\phi
\ll \phi_J$.  Over a finite number of cycles ({\it i.e.}  $n<10^4$ in
Fig.~\ref{Fig23} (b)), transient irreversible dynamics can occur, but
these systems become point-reversible, loop-reversible, or
steady-state irreversible as $n\rightarrow\infty$. Point-reversible
systems tend to form ordered, size-segregated layers, in which
particles cannot collide during simple shear.  Further, the
loop-reversible to irreversible transition in steady-state
$\phi_R(\gamma_{\rm max})$ is bounded in the large-$\gamma_{\rm max}$
limit by the highest packing fraction $\phi_R$ at which systems can
form ordered, size-segregated layers; $\phi_I(\gamma_{\rm max}
\rightarrow \infty) = \phi_R=\pi/4\simeq 0.785$ in 2D and $0.605$ in
3D in the $N \rightarrow \infty$ limit.

In Fig.~\ref{Fig23} (b), scatter plots of $L(n)$ versus $\Delta
r_1(n)$ for 2D systems illustrate the evolution of the dynamics with
increasing $n$.  The points form several well-defined clusters:
point-reversible ($P$) with $L < \tau_L$ and $\Delta r_1 < \tau_r$,
loop-reversible ($L$) with nonzero $L$ ($L > \tau_L$) and $\Delta r_1
< \tau_r$, and irreversible ($I$) with nonzero $L$ ($L > \tau_L$) and
$\Delta r_1$ ($\Delta r_1 > \tau_r$).  The $P$, $L$, and $I$ clusters
are separated by more than $3$ orders of magnitude in $\Delta r_1$ or
$L$.  For region $L$, we also mandate $L>30\Delta r_1$ since systems
with $L<30\Delta r_1$ typically relax to point-reversible states.  We
also enforce $\Delta r_1>0.3\sigma_s$ to define region $I$ since
systems with $\Delta r_1<0.3\sigma_s$ typically relax to point- or
loop-reversible states. Systems that do not fall within the bounds 
defining regions $P$, $L$, and $I$ are categorized as transient
irreversible ($T$).  As $n$ increases, the fraction $F_t$ of systems
in the transient regime vanishes as a power-law
$n^{-\alpha}$ (where $\alpha = 0.56\pm 0.01$), while the fraction of
point-reversible, loop-reversible, and steady-state irreversible
systems saturates near $10^4$ cycles (inset to Fig.~\ref{Fig23} (b)).

\begin{figure}[h]
\begin{center}
\includegraphics[width=0.34\textwidth]{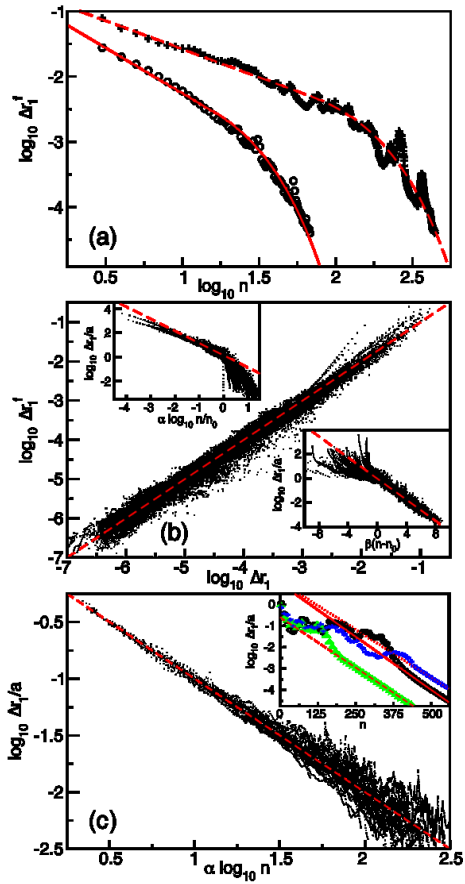}
\caption{(color online) (a) Single cycle mean-square displacement
$\Delta r_1$ versus $n$ for a point-reversible system at $\phi=0.64$
and $\gamma_{\rm max}=0.5$ (circles) and a loop-reversible system at
$\phi=0.8$ and $\gamma_{\rm max}=0.5$ (pluses) with best fits to
$\Delta r_1^f$ (Eq.~\ref{scaling}) indicated by solid and dashed
lines.  (b) Comparison of $\Delta r_1(n)$ (averaged over $16$ initial
conditions for each $\gamma_{\rm max}$ and $\phi$) to $\Delta
r^f_1(n)$ (black dots), where $n$ is the cycle
number, for point-reversible systems in Fig.~\ref{Fig23} with $\Delta
< 0.18$.  $\Delta r_1^f(n)=\Delta r_1(n)$ is indicated by the dashed
line. The top left inset shows $\log_{10} \Delta r_1(n)/a$ versus
$\alpha \log_{10} n/n_c$ (black dots). The dashed line indicates
$\Delta r_1(n)/a = (n/n_c)^{-\alpha}$.  The bottom right inset shows
$\log_{10} \Delta r_1(n)/a$ versus $\beta(n-n_c)$ (black dots).
$\Delta r_1(n)/a = e^{-\beta(n-n_c)}$ is indicated by the dashed
line. (c) $\log_{10} \Delta r_1(n)/a$ versus $\alpha \log_{10} n$
(black dots) for systems in Fig.~\ref{Fig23} that evolve to
loop-reversible dynamical states with $\Delta < 0.04$.  $\Delta
r_1(n)/a= n^{-\alpha}$ is indicated by the dashed line. The inset
shows $\Delta r_1/a$ versus $n$ for three independent initial
conditions (squares, triangles, and pluses) at $\phi=0.76$ and
$\gamma_{\rm max} = 0.8$. Exponential
fits to the large $n$ regime are shown as solid, dashed, and dotted
lines with slopes $\beta=0.029$, $0.026$, and $0.016$, respectively.}
\label{Fig45}
\end{center}
\end{figure}

We also characterized the dynamics of these systems as they
approach steady-state point- and loop-reversible states (Fig.~\ref{Fig45} (a)).
We find that the single cycle
mean-square displacement can be described by a function that
interpolates between power-law and exponential decays at short and
long times, respectively:
\begin{equation}
\Delta r_1^f(n) = f_+(n) (n/n_c)^{-\alpha}+ f_-(n) e^{-\beta(n-n_c)}, 
\label{scaling}
\end{equation}
where $f_{\pm}(n) = (1+e^{ \pm {\overline \gamma} (n-n_c)})^{-1}$, ${\overline \gamma} \sim
1$, $n_c$ is the cycle number at which the decay changes from
power-law to exponential behavior, $\alpha$ is a power-law scaling
exponent, and $\beta$ characterizes the exponential decay.

In Fig.~\ref{Fig45} (b), we plot the best fit $\Delta r^f_1(n)$ versus $\Delta
r_1(n)$ at each $\gamma_{\rm max}$ and $\phi$ (averaged over $16$
initial conditions) for the 2D systems in Fig.~\ref{Fig23} that evolve
to point-reversible states. The scaling function in
Eq.~\ref{scaling} collapses more than $60\%$ of point-reversibile
systems with deviations $\Delta=\langle(\log_{10} \Delta r_1^f(n) -
\log_{10} \Delta r_1(n))^2\rangle < 0.18$.  The top and bottom insets
in Fig.~\ref{Fig45} (b) show the power-law scaling and exponential
decay of $\Delta r_1(n)$ separately.  We find similar scaling for
the approach to loop-reversible states. However, the
exponential decay for loop-reversible systems is difficult to
differentiate from numerical error because the long-time dynamics
occurs at larger $n_c$ and smaller $\Delta r_1$ than that for
point-reversible systems.  In Fig.~\ref{Fig45} (c), we show the
power-law decay for all systems that evolve to
loop-reversible dynamical states.  In the inset, we also show several
systems for which we captured the long-time exponential
decay.

\begin{figure}[h]
\begin{center}
\includegraphics[width=0.4\textwidth]{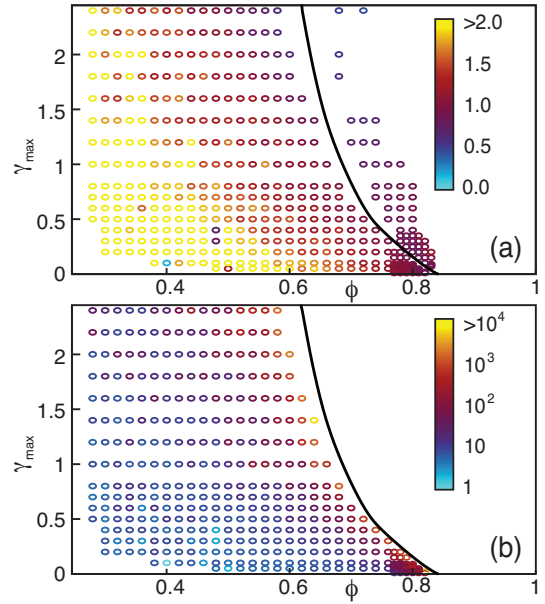}
\caption{(color online) (a) Contour plot of the power-law exponent $\alpha$
(Eq.~\ref{scaling}) versus $\phi$ and $\gamma_{\rm max}$ for
2D systems that evolve to either point- (circles) or loop-reversible
(crosses) states.  (b) Contour plot of the cycle number
$n_c$ (Eq.~\ref{scaling}) that controls the crossover from power-law
to exponential decay versus $\phi$ and $\gamma_{\rm max}$
for the point-reversible systems in (a).}
\label{Fig678}
\end{center}
\end{figure}

In Fig.~\ref{Fig678} (a), we show the power-law scaling exponent
$\alpha$ for 2D systems that evolve to point- and loop-reversible
states versus $\phi$ and $\gamma_{\rm max}$.  We find that $\alpha
\lesssim 1$ for all loop-reversible systems and point-reversible
systems near the crossover from point- to loop-reversible behavior,
which suggests that the origin of the slow dynamics is related to
contact or ``collision'' percolation.  In contrast, $\alpha> 1$ for
point-reversible systems at low $\phi$ and $\gamma_{\rm max}$.  In
Fig.~\ref{Fig678} (b), we plot $n_c$ for point-reversible systems
versus $\phi$ and $\gamma_{\rm max}$. We find that $n_c$ increases
with $\phi$ and $\gamma_{\rm max}$, and appears to be diverging as the
system approaches the transition from point- to loop-reversibility.

To test the stability of steady-state loop-reversible states, we
perturb all particles at strain $\gamma=0$ by an amplitude $\delta$ in
random directions~\cite{supp}.  We then perform cyclic simple shear on
the perturbed system and measure the deviation,
$\Delta_r=\sqrt{(N\sigma_s^2)^{-1}\sum_i\big|\vec{r}^i_{n,0}-\vec{r}^{i,p}_{n,0}\big|^2}$,
where $\vec{r}^{i,p}_{n,0}$ are the coordinates of the perturbed
system after $t$ cycles required to reach steady steady at each $\phi$
and $\gamma_{\rm max}$.  We find that loop-reversible systems are
stable ($\Delta_r<\tau_r$) for perturbations $\delta < \delta_c \simeq
10^{-1}$, {\it i.e.} perturbations as large as one-tenth of a particle
diameter, where $\delta_c$ is relatively insensitive to $\phi$ for
$\gamma_{\rm max} \lesssim 1$.

In conclusion, we studied the extent to which particle-scale motion
can be reversible in athermal systems undergoing cyclic quasistatic
loading.  We identified two types of reversible behavior.  For
point-reversible states, particles do not collide and
therefore trivially retrace their paths.  For loop-reversible
states, all particles undergo multiple collisions and have
complex trajectories, yet all particles return to the locations they
were in at the beginning of the cycle.  We determined the regions of
packing fraction and strain amplitudes in 2D and 3D where these
dynamical states are stable.  In particular, we find that
loop-reversible states occur over a range of packing fraction from
contact percolation~\cite{shen} to jamming onset, and thus our results
emphasize that complex spatiotemporal dynamics are found well below
$\phi_J$.  Loop-reversible dynamical states represent a novel
form of self-organization that will enable reliable preparation of
configurations with particular structural and rheological properties
over a broad range of packing fractions.

We acknowledge support from NSF grant numbers NSF MRSEC DMR-1119826
(CS), DMR-1006537 (RH), CBET-0968013 (MS), and CBET-0967262 (CO).  We
also thank S. Papanikolaou, S. S. Ashwin, and T. Bertrand for helpful
discussions.


\begin{thebibliography}{99}

\bibitem{schall}
P. Schall, D. A. Weitz, and F. Spaepen, {\it Science} {\bf 318} (2007) 1895. 

\bibitem{bmg}
J. S. Harmon, M. D. Demetriou, W. L. Johnson, and K. Samwer, {\it Phys. 
Rev. Lett.} {\bf 99} (2007) 135502.  

\bibitem{stz}
M. L. Falk and J. S. Langer, {Ann. Rev. Condens. Matt. Physics} {\bf 2} (2011)
353. 

\bibitem{nicolas} M. Nicolas, P. Duru, and O. Pouliquen, {\it
Eur. Phys. J. E} {\bf 3} (2000) 309.

\bibitem{nowak}
E. R. Nowak, J. B. Knight, E. Ben-Naim, H. M. Jaeger, and S. R. Nagel, 
{\it Phys. Rev. E} {\bf 57} (1998) 1971. 

\bibitem{Lundberg2008} M. Lundberg, K. Krishan, N. Xu, C. S. O'Hern, 
and M. Dennin, {\it Phys. Rev. E} {\bf 77} (2008) 041505.

\bibitem{Pine2005} D. J. Pine, J. P. Gollub, J. F. Brady, 
and A. M. Leshansky, {\it Nature} {\bf 438} (2005) 997.

\bibitem{Corte2008} Laurent Corte, P. M. Chaikin, J. P. Gollub, and
D. J. Pine, {\it Nature Physics} {\bf 4} (2008) 420. 

\bibitem{slotterback}
S. Slotterback, M. Mailman, K. Ronaszegi, M. van Hecke, M. Girvan, and W. 
Losert, {\it Phys. Rev. E} {\bf 85} (2012) 021309. 

\bibitem{shen}
T. Shen, C. S. O'Hern, and M. D. Shattuck, {\it Phys. Rev. E} {\bf 85} (2012) 
011308.  

\bibitem{allen} M. P. Allen and D. J. Tildesley, {\it Computer 
Simulation of Liquids} (Oxford University Press, New York, 1987).

\bibitem{gao}
G.-J. Gao, J. Blawzdziewicz, and C. S. O'Hern, {\it Phys. Rev. E} {\bf 80} 
(2009) 061303. 

\bibitem{xu}
N. Xu and C. S. O'Hern, {\it Phys. Rev. Lett.} {\bf 94} (2005) 
055701. 

\bibitem{supp} See Supplementary Materials, which include results for
2D and 3D systems as a function of system size $N$, shear
strain increment $\Delta \gamma$, and energy minimization threshold
$V_{\rm th}$.

\end{thebibliography}
\end{document}